# Co-Phased 360-Degree Profilometry of Discontinuous Solids with 2-Projectors and 1-Camera

**Manuel Servin, Guillermo Garnica and J. M. Padilla.**

*Centro de Investigaciones en Optica A. C., 115 Loma del Bosque, 37150 Leon Guanajuato, Mexico*
*mservin@cio.mx , garnica@cio.mx , moises@cio.mx .*

**Abstract:** Here we describe a co-phased 360-degree fringe-projection profilometer which uses 2-projectors and 1-camera and can digitize discontinuous solids with diffuse light surface. This is called co-phased because the two phase demodulated analytic-signals from each projection are added coherently. This 360-degree co-phased profilometer solves the self-generated shadows cast by the object discontinuities due to the angle between the camera and the single white-light fringe projector in standard profilometry.

## 1. Introduction

One sided or 180-degree fringe projection profilometry using phase-demodulation techniques is well known since the classical paper by Takeda et al. in 1982 [1]. However to obtain a full 360-degree profilometer one needs to position the solid over a turntable to have access to all solid perspectives. Previous efforts have mainly concentrated in digitizing very smooth, quasi-cylindrical objects [2]. This is because smooth, quasi-cylindrical solids are easier to digitize with previous 360-degree profilometers. Here we are proposing a co-phased 2-projector, 1-camera 360-degree profilometer which is capable to digitize highly discontinuous 3D solids. This co-phased 2-projectors, 1-camera profilometer is a blending of two previous published techniques by the same authors [3,4]. We have digitized a human skull model all around it (360-degrees). The digitizing of this plastic skull shows the 3D digitizing capabilities of this profilometer to digitize discontinuous non-convex 3D solids.

## 2. Co-phased 180-degree fringe-projection profilometry using 2-projections and 1-camera

The left-*L* and right-*R* projected fringes as seen by the CCD camera are given by,

$$I_L(x,y,\alpha) = a_L(x,y) + b_L(x,y)\cos[\omega_0 x - g\,z(x,y) + \alpha]; \quad \omega_0 = v_0 \cos(\varphi_0),\ g = \tan(\varphi_0),$$
$$I_R(x,y,\alpha) = a_R(x,y) + b_R(x,y)\cos[\omega_0 x + g\,z(x,y) + \alpha]; \quad \alpha \in \{0, \pi/2, \pi, 3\pi/2\}. \tag{1}$$

Where the digitizing object is $z = z(x,y)$ and $\alpha \in \{0, \pi/2, \pi, 3\pi/2\}$ are the 4 phase-steps used.

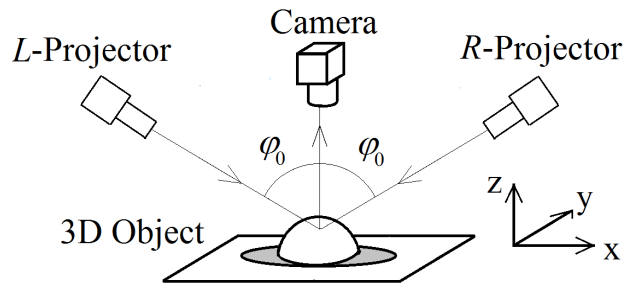

Figure 1. Co-phased 180-degree fringe-projection profilometer with 2-projectors and 1-camera [3].

We then use a 4-step phase-shifting demodulation algorithm for both projected images as,

$$A_L(x,y)e^{igz(x,y)} = I_L(0) + I_L(\pi/2)e^{i\pi/2} + I_L(\pi)e^{i\pi} + I_L(3\pi/2)e^{i3\pi/2},$$
$$A_R(x,y)e^{-igz(x,y)} = I_R(0) + I_R(\pi/2)e^{i\pi/2} + I_R(\pi)e^{i\pi} + I_R(3\pi/2)e^{i3\pi/2}. \tag{2}$$

Note that the analytic signal of the right projector $A_R(x,y)e^{-igz(x,y)}$ has negative recovered phase with respect to the left projector $A_L(x,y)e^{igz(x,y)}$. These two analytic signals in Eq. (2) have complementary shadows and their digital co-phased sum is,

$$A_{L+R}(x,y)e^{i g z(x,y)} = A_L(x,y)e^{i k g(x,y)} + \left[A_R(x,y)e^{-i g z(x,y)}\right]^*,$$
$$A_{L+R}(x,y)e^{i g z(x,y)} = \left[A_L(x,y) + A_R(x,y)\right]e^{i g z(x,y)}.$$
(3)

The analytic signal $A_{L+R}(x,y)\exp[i g z(x,y)]$ therefore has no self-shadows cast by $z(x,y)$ as seen by the camera.

### 3. Co-phased 360-degree fringe projection profilometry with 2-projectors and 1-camera

The co-phased 2-projectors 1-camera, 360-degree ($\varphi \in [0, 2\pi)$) profilometer set-up is shown in Fig. 2,

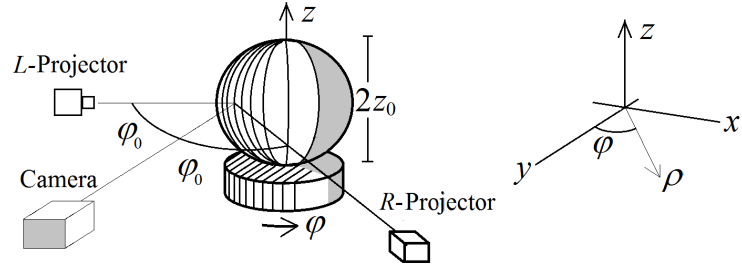

Fig. 2. Co-phased 360-degree profilometer with 2-projectors and 1-camera to digitize discontinuous objects $\rho=\rho(z,\varphi)$.

For each rotation step $\varphi$ we grab the central column CCD pixels of the solid $\rho = \rho(z,\varphi)$, $z \in [-z_0, z_0]$, $\varphi \in [0, 2\pi)$. We then use a 4-step phase-shifting algorithm to phase-demodulate their analytic signals,

$$A_L(z,\varphi)e^{i g \rho(z,\varphi)} = I_L(0) + I_L(\pi/2)e^{i \pi/2} + I_L(\pi)e^{i \pi} + I_L(3\pi/2)e^{i 3\pi/2},$$
$$A_R(z,\varphi)e^{-i g \rho(z,\varphi)} = I_R(0) + I_R(\pi/2)e^{i \pi/2} + I_R(\pi)e^{i \pi} + I_R(3\pi/2)e^{i 3\pi/2}.$$
(4)

Their digital co-phased sum is defined everywhere because they have complementary shadows,

$$A_{L+R}(z,\varphi)e^{i g \rho(z,\varphi)} = A_L(z,\varphi)e^{i g \rho(x,y)} + \left[A_R(z,\varphi)e^{-i g \rho(z,\varphi)}\right]^*,$$
$$A_{L+R}(z,\varphi)e^{i g \rho(z,\varphi)} = \left[A_L(z,\varphi) + A_R(z,\varphi)\right]e^{i g \rho(x,y)}; \qquad z \in [-z_0, z_0], \varphi \in [0, 2\pi].$$
(5)

### 4. Experimental results

Fig. 3 show the images from the left and the right projectors; generating left and right shadows at the discontinuities

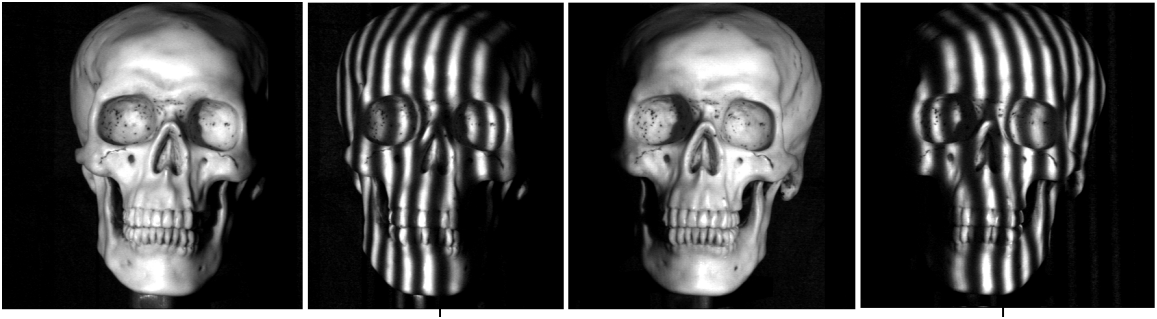

Fig. 3. The digitizing plastic skull from the left and right illumination perspectives

The two 4-step phase-shifted algorithms used to demodulated the 2 analytic signal perspectives are,

$$A_L(z,\varphi)e^{i g\, Skull(z,\varphi)} = I_L(0) + I_L(\pi/2)e^{i \pi/2} + I_L(\pi)e^{i \pi} + I_L(3\pi/2)e^{i 3\pi/2},$$
$$A_R(z,\varphi)e^{-i g\, Skull(z,\varphi)} = I_R(0) + I_R(\pi/2)e^{i \pi/2} + I_R(\pi)e^{i \pi} + I_R(3\pi/2)e^{i 3\pi/2}.$$
(6)

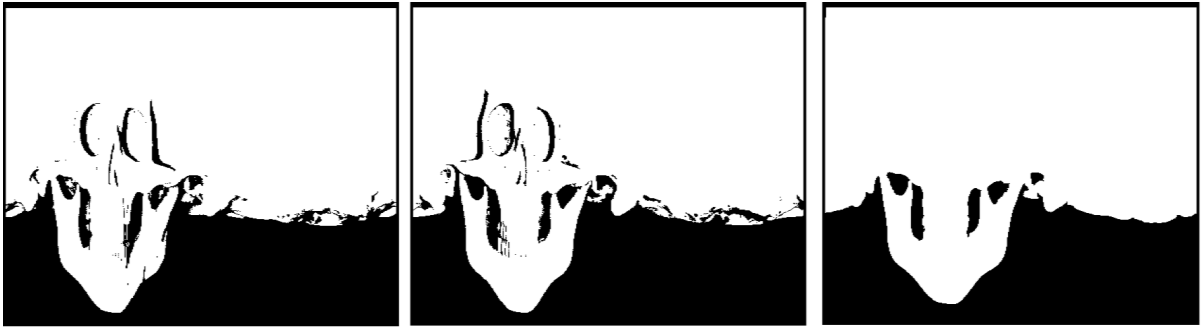

Fig. 4. Binary masks of high amplitude contrast for the left, the right and the sum of the analytic signals in Eqs. (6) and (7).

Now let us sum the two co-phased analytic signals from the left and the right projections in Eq. (6).

$$A_{L+R}(z,\varphi)e^{ig\,Skull(z,\varphi)} = A_L(z,\varphi)e^{ig\,Skull(x,y)} + \left[A_R(z,\varphi)e^{-ig\,Skull(z,\varphi)}\right]^*,$$
$$A_{L+R}(z,\varphi)e^{ig\,Skull(z,\varphi)} = \left[A_L(z,\varphi)+A_R(z,\varphi)\right]e^{ig\,Skull(x,y)}; \qquad z \in [-z_0, z_0],\ \varphi \in [0, 2\pi]. \qquad (7)$$

The two analytic signals in Eq. (6) have complementary shadows at the skull discontinuities, but their co-phased sum $A_{L+R}(x,y)\exp[i\,g\,\rho(z,\varphi)]$ in Eq. (7) has no one (see Fig. 4).

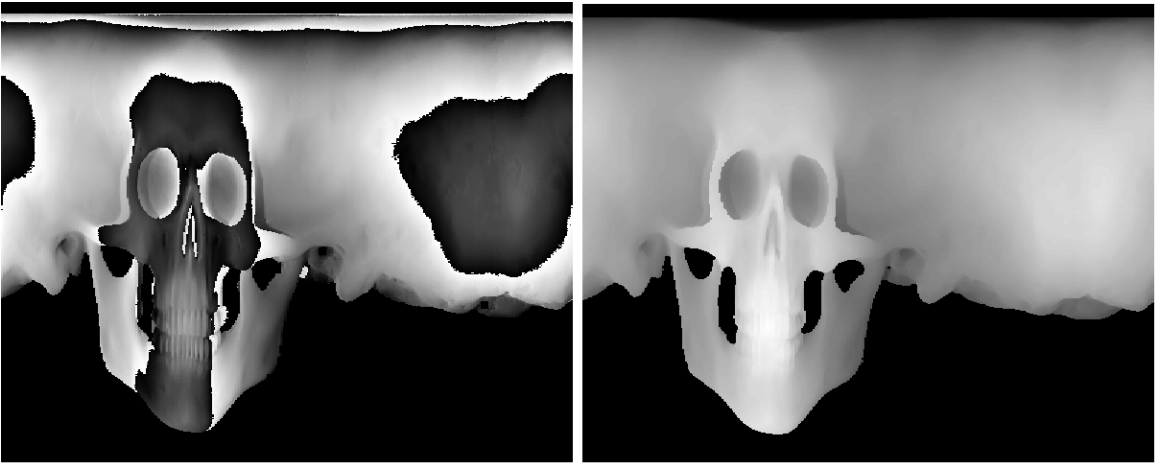

Fig. 5. The left panel shows the wrapped phase of the digitized skull and the right panel shows the corresponding unwrapped phase of the co-phased sum in Eq. (7); both in flat cylindrical-coordinates.

Figure 5 shows the wrapped phase (left panel) and the unwrapped phase (right panel) of the co-phased sum of the two projections using the proposed 360-degree profilometer. Figure 5 shows these phases in two-dimensional (2D) cylindrical coordinates. Finally Fig. 6 shows two 3D-perspectives in the 3D-Euclidian space of the digitized plastic skull formed from the flat unwrapped phase shown in Fig. 5.

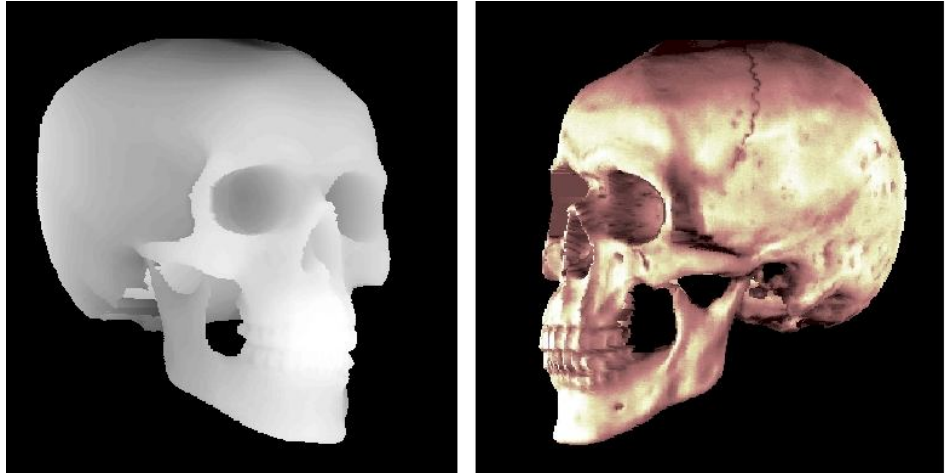

Fig. 6. The left panel shows the 3-dimensional rendering of the plastic skull in gray-level phase-values. The right panel shows another skull perspective with the magnitude of the fringes superimposed.

## 5.- Conclusions

We have presented a white-light fringe-projection 360-degree profilometer which is capable of solving the self-generated shadows cast by highly discontinuous solids. To achieve this, the profilometer must have two fringe projectors at both sides of the single camera that grabs the fringe patterns. Also we have used the amplitude of the phase-demodulated analytic-signal of the co-phased sum to mask the region where the 3D object resides in the flat cylindrical coordinate representation.